\documentclass[3p,sort&compress]{elsarticle}
\usepackage{mathpazo}
\usepackage{amssymb}
\usepackage{longtable}
\usepackage{graphics}
\usepackage{graphicx}
\usepackage[italian,english]{babel}
\usepackage[bf, small, hang]{caption}
\usepackage[bf, small, hang]{subfigure}
\usepackage{amsmath}
\usepackage[colorlinks]{hyperref}
\usepackage{booktabs}

\let\today\relax
\makeatletter
\def\ps@pprintTitle{%
\let\@oddhead\@empty
\let\@evenhead\@empty
\def\@oddfoot{}%
\let\@evenfoot\@oddfoot}
\makeatother

\begin{document}

\title{On the appearance of fractional operators in non-linear stress-strain relation of metals} 
	\author[cor2]{F. P. Pinnola\corref{fn1}}
	\ead{francesco.pinnola@unisalento.it}
	\author[cor1]{G. Zavarise}
	\ead{giorgio.zavarise@polito.it}
	\author[cor2]{A. Del Prete}
	\ead{antonio.delprete@unisalento.it}	
	\author[cor2]{R. Franchi}
	\ead{rodolfo.franchi@unisalento.it}

	\cortext[fn1]{Corresponding author at: Department of Innovation Engineering, Universit\`a del Salento, Ed. La Stecca, SP 6 Lecce-Monteroni, 70100, Lecce, Italy}

	\address[cor2]{Department of Innovation Engineering, Universit\`a del Salento, Ed. La Stecca, SP 6 Lecce-Monteroni, 70100, Lecce, Italy}
	\address[cor1]{Department of Structural, Building and Geotechnical Engineering, Politecnico di Torino, C.so Duca degli Abruzzi 24, 10129 Torino, Italy}

\begin{abstract} 
Finding an accurate stress-strain relation, able to describe the mechanical behavior of metals during {forming} and machining processes, is an important challenge in several fields of mechanics, with significant repercussions in the technological field. Indeed, in order to predict the real mechanical behavior of materials, constitutive laws must be able to take into account elastic, viscous and plastic phenomena. Most constitutive models are based on empirical evidence and/or theoretical approaches, and provide a good prediction of the mechanical behavior of several materials. 

Here we present a non linear stress-strain relation based on fractional operators. The proposed constitutive law is based on integral formulation, and takes into account the viscoelastic behavior of the material and the inelastic phenomenon that appears when the stress reaches a particular yielding value. A specific case of the proposed constitutive law for imposed strain history is used to fit experimental data obtained from tensile tests on two kind of metal alloys. A best-fitting procedure demonstrates the accuracy of the proposed stress-strain relation and its results are  compared to those obtained with some classical models. We conclude that the proposed model provides the best results in predicting the mechanical behavior for low and high values of stress/strain.

\end{abstract}

\begin{keyword}{
Fractional calculus, viscoelastoplasticity, rate-dependent model, experimental tests}
\end{keyword}

\maketitle

%\section*{Notation}

\section{Introduction}
{\noindent High performances materials such as steels and aluminum alloys are extensively used in all the applications where advanced mechanical properties and good strength/weight ratio are required. Other kinds of materials, like titanium and nickel alloys and superalloys, are applied in all the cases requiring long-time strength and toughness at high temperature and/or good creep resistance. A large number of mechanical metallic components (made using the above-mentioned materials) are manufactured through forming and machining processes. These continue to dominate among all manufacturing processes, despite additive processes allowing to obtain directly products close to the final shape. Forming makes use of compressive forces and is considered to be better than casting, thanks to its production of parts with denser microstructures, better grain patterns and less porosity. Moreover, machining permits to obtain a good surface integrity (roughness, residual stresses and microstructure) that allows the use of the worked components in all the applications in which high fatigue strength is required. In this context, a large number of works have been carried out to investigate and to optimize forming and machining processes of metal alloys, in order to improve the quality of the components and, at the same time, to increase the productivity and lower the cost.
Several improvements in terms of better understanding of forming and machining processes have been also obtained in the last decades, thanks to the heavy use of advanced mechanical models of the stress-strain relation and numerical simulations.

Nowadays, numerical simulations are crucial to investigate the behavior of materials under several mechanical conditions, with evident advantages in terms of time and cost saving during the process design phase \cite{1,2}. Among these simulations, finite element simulation permits to study some aspects of the process that are difficult to investigate experimentally. However, to set up a model having a good prediction capability, the development of a reliable stress-strain relation of the considered material is the fundamental basis of a numerical investigation \cite{3,4}. In particular, in order to predict the mechanical behavior in the numerical simulations, the choice of a proper model able to describe the stress-strain relation of the material during the manufacturing process is a fundamental step. 
Obviously, in the context of material processing, the simple elastic stress-strain relation is not sufficiently representative of the real mechanical behavior. Therefore, it is important to find an accurate mechanical model that takes into account elastic, viscous and plastic behaviors. 
}

In mechanics, usually, the stress-strain models are based on empirical and/or analytical approaches. In the literature there are several empirical models able to describe the elasto-platic phenomenon. Probably the most used are the Hollomon (H) and Ramberg-Osgood (RO) models \cite{Mandelson,Malvern,Lubliner,Simo,Han-Chin}. In particular, according to the H model, the stress $\sigma$ and strain $\varepsilon$ are related by a power-law function as follows,
\begin{equation}
\sigma=K_H\varepsilon^{n_H}
\label{eq1}
\end{equation}
where $K_H$ and $n_H$ are the material parameters \cite{Hollomon}. In the RO model the stress-strain relation is given as
\begin{equation}
\varepsilon=\frac{\sigma}{E}+\left(\frac{\sigma}{H_{RO}}\right)^{n_{RO}}
\label{eq2}
\end{equation}
where ${E}$ is the elastic modulus and $H_{RO}$ and $n_{RO}$ are two material parameters \cite{Nadai,Ramberg-Osgood}. Both models are used to describe the stress-strain behavior of metals during several experimental tests \cite{exp0,exp1,exp2,exp3,exp4,exp5}. Other models used to perform the characterization of the stress-strain relation of elastoplastic material can be found in \cite{Ludwigson,Swift,Ludwick}.

On the other hand, there are several theories and related mechanical models based on analytical approaches \cite{Mandelson,Malvern,Lubliner,Simo,Han-Chin,Caddemi}. Probably the most commonly used theory is the so-called \emph{flow theory}, or \emph{incremental theory of plasticity}. The theory considers an infinitely slow process and regards plastic materials as inviscid. Usually, it defines a yielding surface that denotes a change in the mechanical behavior of the material. In particular, after a linear elastic range bounded by a limit value of yielding stress $\sigma_Y$, another kind of mechanical behavior takes place, in which the deformation is a linear combination of the elastic and plastic parts, i.e., $d\varepsilon = d\varepsilon^e + d\varepsilon^p$. The plastic part cannot be recovered, while the elastic part is fully recoverable. This theory is due to the works of several scientists, e.g. Melan, Prager, Hodge, Hill, Drucker, Budiansky, Koiter, etc., and in its early form it does not take into account the rate effect. In fact, it is also known as \emph{rate-independent plasticity}, since both strain-rate and stress-rate do not influence the constitutive law. Other approaches have provided an accurate description of the mechanical behavior of material taking into account also the stress and or the strain-rate. This kind of rate-dependent theory is known as \emph{viscoplasticity} \cite{Mandelson,Malvern,Lubliner,Simo,Han-Chin,Cernocky,Auricchio,Auricchio-Taylor,Deseri}.

A different way to describe a non linear stress-strain relation has been developed by Iliushin \cite{Lubliner,Han-Chin}, who provides a formulation similar to the Boltzmann integral formulation used in viscoelasticity \cite{Flugge,Christensen}. In this context, a particular stress-strain relation has been introduced by Valanis \cite{Val1,Val2,Val3,Val4,Val5} in his endochronic theory of viscoplasticity. Such theory does not define a yielding surface, but introduces an intrinsic time scale which is monotonically increasing (endochronic time). Originally, Valanis' theory has been developed to describe the mechanical behavior of metals, but its field of application has been extended to other materials \cite{Bazant}.
In this regards, we note the endochronic stress-strain relation for isotropic and plastically incompressible materials obtained by Peng and Porter \cite{PengPorter1}
\begin{equation}
\sigma_{jl}(t)=K \varepsilon_{kk}^e(t)\delta_{jl}+\int_{0}^{t}\rho\left(z(t)-z(\tau)\right)\dot\varepsilon_{jl}^i(\tau)d\tau
\label{eqendo}
\end{equation}
where $K$ is the elastic Bulk modulus, $\delta_{jl}$ is the Kronecker delta, $\varepsilon_{kk}^e(t)$ and $\varepsilon_{jl}^i(t)$ are the volumetric and the deviatoric components of the strain (the apex $e$ and $i$ stand for elastic and inelastic, respectively), the kernel $\rho(z)$ is a memory function known as \emph{pseudo-relaxation function} \cite{Lubliner}, $z$ is a function both of time and inelastic deformation, and it is called intrinsic (or \emph{endochronic}) time scale. Eq.~(\ref{eqendo}) shows that the inelastic stress-strain relation can be modeled in a similar way to the Boltzmann superposition integral which is often used in viscoelasticity \cite{Malvern,Flugge,Christensen}. Indeed, the stress-strain relation for viscoelastic isotropic and linear elastic incompressible material is 
\begin{equation}
\sigma_{jl}(t)=K \varepsilon^e_{kk}(t)\delta_{jl}+\int_{0}^{t}R(t-\tau)\dot\varepsilon^{ve}_{jl}(\tau)d\tau
\label{eqvisco}
\end{equation}
where the apex $ve$ stands for viscoelastic, $R(t)$ is the \emph{relaxation function} that can be a series of exponentials \cite{Flugge,Christensen} or a power-law function of time \cite{Koeller,Makris,Alotta}. Observe that Eq.~(\ref{eqendo}) and Eq.~(\ref{eqvisco}) describe different phenomena with the same mathematical operators, that is, the hereditary integral with different kernel. The substantial difference between the two integral kernels lies in the involved variable, i.e. time $t$ for viscoelasticity, and intrinsic time $z$ for viscoplasticity. The first one is an independent variable whereas the latter is a function of the time and the deformation.

In this paper we present a new uniaxial stress-strain relation based on integral formulation. The proposed constitutive law allows to take into account the viscoelastic behavior of the material and the inelastic properties that appear when the stress reaches a particular yielding value and the irreversible phenomenon onsets. 

{The manuscript is organized as follow. Section~\ref{sect2} introduces the proposed uniaxial model of viscoplastic behavior in terms of non-linear stress-strain relation. It contains the related hypotheses and shows how, for the considered cases, a non-linear behavior can be described as a summation of two convolution integrals. In Section~\ref{sect3} the proposed stress-strain relation is used to describe the mechanical behavior during a tensile test. Considering the results from this kind of test, Section~\ref{sect4} contains the best-fitting of experimental data performed with the aid of the proposed model. Finally, in Section~\ref{sect5} concluding remarks and some comments about the advantages of the proposed model with respect to other ones available in literature are reported.}

\section{Uniaxial stress-strain relation}
\label{sect2}
\noindent In this section a non-linear constitutive model in terms of stress-strain relation is proposed. In this regards we consider that stress and strain are time-dependent uniaxial fields. Therefore, time evolution has to be considered both for strain $\varepsilon(t)$ and stress $\sigma(t)$ histories. In this way both strain-rate and stress-rate influence the constitutive relation. The model proposed below is a non-linear stress-strain relation. However, under some physical/mathematical restrictions the model leads to a summation of two linear operators. 
In particular, assume that
\begin{itemize}
\item Strain deformation $\varepsilon(t)$ is a positive monotonic increasing function of time, i.e. if for all $t_i$ and $t_j$ such that $t_i\leqslant t_j$, one has 
\begin{equation}
0\leqslant\varepsilon(t_i)\leqslant\varepsilon(t_j).
\end{equation}
\item Strain deformation for all values of $t$ is a summation of viscoelastic and inelastic deformation. That is,
\begin{equation}
\varepsilon(t)=\varepsilon^{ve}(t)+\varepsilon^i(t),
\label{eq6b}
\end{equation}
where $\varepsilon^{ve}(t)$ is the viscoelastic part and $\varepsilon^i(t)$ denotes the inelastic one.
\item Inelastic deformation is unlimited, whereas the viscoelastic one is bounded to a maximum value, $\varepsilon_Y$, that corresponds to the yield limit $\varepsilon_Y=f(\sigma_Y)$. The time when the viscoelastic deformation reaches the maximum limit is $t_Y$, then $\varepsilon^{ve}(t_Y)=\varepsilon_Y$. Inelastic deformation $\varepsilon^i(t)$ increases only if the viscoelastic deformation reaches the limit value $\varepsilon_Y$, and then $\varepsilon^i(t)>0$, $\forall \, t:\,t> t_Y$. Therefore,
\begin{equation}
\varepsilon(t)=
\begin{cases}
\varepsilon^{ve}(t)&\mathrm{for}\; 0<t\leqslant t_Y \\
\varepsilon_Y+\varepsilon^i(t)&\mathrm{for}\; t>t_Y. 
\end{cases}
\label{eq6}
\end{equation}
The inelastic deformation onsets from the yielding point $P_Y=\{\varepsilon(t_Y),\sigma(t_Y)\}$. Observe that also the time of yielding $t_Y$ is function of the deformation $\varepsilon_Y$, thus $t_Y=f(\varepsilon_Y)=f(\sigma_Y)$.
\item Strain deformation increases during the time, therefore $\frac{d}{dt}\varepsilon(t)=\dot\varepsilon(t)>0$ $\forall \, t:\,t>0$. Under this assumption and taking into account Eq.s~(\ref{eq6b}) and (\ref{eq6}), the following relation holds true
\begin{equation}
\dot\varepsilon(t)=\dot\varepsilon^{ve}(t)+\dot\varepsilon^{i}(t)=
\begin{cases}
\dot\varepsilon^{ve}(t)&\mathrm{for}\; 0<t\leqslant t_Y, \\
\dot\varepsilon^i(t)&\mathrm{for}\; t>t_Y.
\end{cases}
\label{eq6c}
\end{equation}
\end{itemize}

By using an integral formulation of the stress-strain relation, for a virgin material at initial time $t=0$, the stress history can be expressed by the following convolution integral
\begin{equation}
\begin{split}
\sigma(t,\varepsilon^i)&=\int_0^tR\left(t-\tau\right)d\varepsilon^{ve}(\tau)+\int_0^t\rho\left(t-\tau,\varepsilon^i\right)d\varepsilon^{i}(\tau)\\
&=\int_0^tR\left(t-\tau\right)\dot\varepsilon^{ve}(\tau) d\tau+\int_{t_Y}^t\rho\left(t-\tau,\varepsilon^i\right)\dot\varepsilon^{i}(\tau) d\tau
\end{split}
\label{eq7}
\end{equation}
where the first integral kernel $R(t)$ is the relaxation modulus used in viscoelastic theory and reported in Eq.~(\ref{eqvisco}), while the second kernel $\rho\left(t,\varepsilon^i\right)$ is function of time $t$ and inelastic deformation $\varepsilon^i$, and results similar to the pseudo-relaxation modulus in Eq.~(\ref{eqendo}). If the viscoelastic deformation does not reach the limit bound $\varepsilon_Y$, the inelastic deformation does not arise and the relation in Eq.~(\ref{eq7}) reverts to the classical Boltzmann superposition integral used in linear viscoelasticity. In Eq.~(\ref{eq7}) the first integral considers the increment of stress history due to the linear viscoelastic effect, whereas the second one is related to the time-evolution of inelastic deformation. Other works assume that after the yield stress $\sigma_Y$, and the related yield strain $\varepsilon_Y$, the mechanical properties of the material changes also for the viscoelastic properties \cite{Yu,Thibaud,Halilovic,Mendiguren,Dimino,Meng}. This fact implies that the relaxation modulus can change after the time $t_Y$, however this possibility in the presented formulation is not contemplated. 

Obviously, the two kernels $R(t)$ and $\rho(t,\varepsilon^i)$ are not the same, since they are related to two different kind of deformation. In linear viscoelasticity several experimental investigations have shown that the relaxation functions of several materials are proportional to a power-law function of time \cite{Nutting,Gemant,Nutting2,Valenza1,Valenza2,Alotta1,Spanos,Alotta2,rampa}. {These works have shown that a power-law kernel is able to describe several experimental evidences. Thanks to this capability,} assume that both moduli are power-law functions of the time. That is,
\begin{equation}
R(t)=A t^{-\alpha},\;\;\;\;\rho(t,\varepsilon^i)=B t^{-\beta}U\left(|\dot\varepsilon^i|-|\dot\varepsilon^{ve}|\right),
\label{eq9a}
\end{equation} 
where four parameters $A$, $\alpha$, $B$ and $\beta$ are involved, $U(\cdot)$ denotes the unit step function, that is,
\begin{equation}
U(x)=\left\{\begin{split}1,\;\;x\geqslant0,\\
0,\;\;x<0.
\end{split}\right.
\label{eq17}
\end{equation}
where is $x$ is an independent variable.

The unit step in Eq.~(\ref{eq9a}) is introduced to take into account the viscoelastic-back during the unloading process. However, under the aforementioned assumptions that the deformation is a positive monotonic increasing function of time, i.e. $\varepsilon(t)>0$ and $\varepsilon^i(t)>\dot\varepsilon^{ve}(t)$ for all $t>t_Y$, the moduli in Eq.~(\ref{eq9a}) can be rewritten as
\begin{equation}
R(t)=A t^{-\alpha},\;\;\;\;\rho(t)=B t^{-\beta},
\label{eq9}
\end{equation} 
where $A$ and $B$ are known as anomalous moduli \cite{Valenza1,Valenza2}, since their dimension is $\mathrm{Pa}\,\mathrm{sec}^\alpha$ and $\mathrm{Pa}\, \mathrm{sec}^\beta$, respectively.
Placing the selected power-law kernels of Eq.~(\ref{eq9}) into the Eq.~(\ref{eq7}), the stress-strain relation becomes
\begin{equation}
\sigma(t)=\tilde{A}\Big({}_CD_{0^+}^{\alpha}\varepsilon^{ve}\Big)(t)+\tilde{B}\left({}_CD_{t_Y^+}^{\beta}\varepsilon^i\right)(t),
\label{eq10}
\end{equation}
where $\left({}_CD_{a^+}^{\gamma}\cdot\right)(t)$ denotes the $\gamma$-order Caputo's fractional derivative with lower bound $a$ \cite{Oldham,Samko,Miller,Pod,Kilbas}. That is,  
\begin{equation}
\Big({}_CD^\gamma_{a^+}f\Big)(t):=\left\{\begin{split}
&\frac{1}{\Gamma(n-\gamma)}\int_a^t {\frac{f^{(n)}(\tau)}{(t-\tau)^{n-1+\gamma}}d\tau}, &n-1<\gamma<n,\,\gamma\in\mathbb{R},\\
&\frac{d^n}{dt}f(t), &\gamma=n,\,n\in\mathbb{N}.
\end{split}\right.
\label{eq11}
\end{equation} 
Taking into account the Caputo's fractional derivative in Eq.~(\ref{eq11}) and the selected moduli in Eq.~(\ref{eq9}), the involved parameters in Eq.~(\ref{eq10}) becomes
\begin{equation}
A=\frac{\tilde A}{\Gamma(1-\alpha)},\;\;\;B=\frac{\tilde{B}}{\Gamma(n-\beta)},
\label{eq12}
\end{equation} 
being $\Gamma(\cdot)$ the Euler gamma function, defined as
\begin{equation}
\Gamma(x)=\int_0^\infty e^{-t}t^{x-1}dt.
\label{eq13}
\end{equation} 
In Eq.~(\ref{eq12}) it is assumed that the fractional orders $\alpha$ is less than one. This is always true because the physical viscoelastic phenomenon implies this limitation, while for the order $\beta$ the upper limit is not defined a priori.

Observe that, according to the definition of Caputo's fractional derivative in Eq.~(\ref{eq11}), the first term of Eq.~(\ref{eq10}) represents the stress-strain relation commonly used in linear fractional viscoelasticity. The elastic Hookean relation and the viscous Newtonian behavior are contained in the first convolution integral for the limit values of order $\alpha$. In particular, for $\alpha=0$, the viscoelastic deformation becomes elastic $\varepsilon^{ve}(t)=\varepsilon^e(t)$, and Eq.~(\ref{eq10}) becomes 
\begin{equation}
\sigma(t)=\tilde{A}\varepsilon^{e}(t)+\tilde{B}\left({}_CD_{t_Y^+}^{\beta}\varepsilon^i\right)(t),
\label{eq14}
\end{equation}
where $\tilde{A}=A=E$ is nothing else than a Young's modulus. Whereas, if $\alpha=1$, the deformation in the first fractional operator becomes a purely viscous deformation $\varepsilon^{ve}(t)=\varepsilon^v(t)$, and Eq.~(\ref{eq10}) yields
\begin{equation}
\sigma(t)=\tilde{A}{\dot\varepsilon^{v}}(t)+\tilde{B}\left({}_CD_{t_Y^+}^{\beta}\varepsilon^i\right)(t),
\label{eq15}
\end{equation}
where the coefficient $\tilde{A}=A=\mu$ becomes a viscosity. 

When $\beta$ is in the range $1\leqslant\beta\leqslant2$ Eq.~(\ref{eq15}) is similar to the stress-strain relation of non-Newtonian fluid proposed by Yin et al. in \cite{Yin}.

\section{Stress-strain relation for tensile test}
\label{sect3}
\noindent Several experimental investigations have been obtained by imposing a ramp as strain history during the displacement control tensile test. In this section, considering this kind of experiment, some known cases are modeled. Moreover, it is shown that some idealized stress-strain relations can be seen as particular case of the proposed model in Eq.~(\ref{eq10}), if the imposed strain history is a ramp. 

During a displacement control test, the imposed deformation history increases constantly during the time for $t>0$. Therefore,
\begin{equation}
\varepsilon(t)=\dot\varepsilon_0\,t\,U(t),
\label{eq16}
\end{equation}
where $\dot\varepsilon_0$ is the initial deformation rate. 

\begin{figure}[h]
\begin{center}
\includegraphics[width=0.475\textwidth]{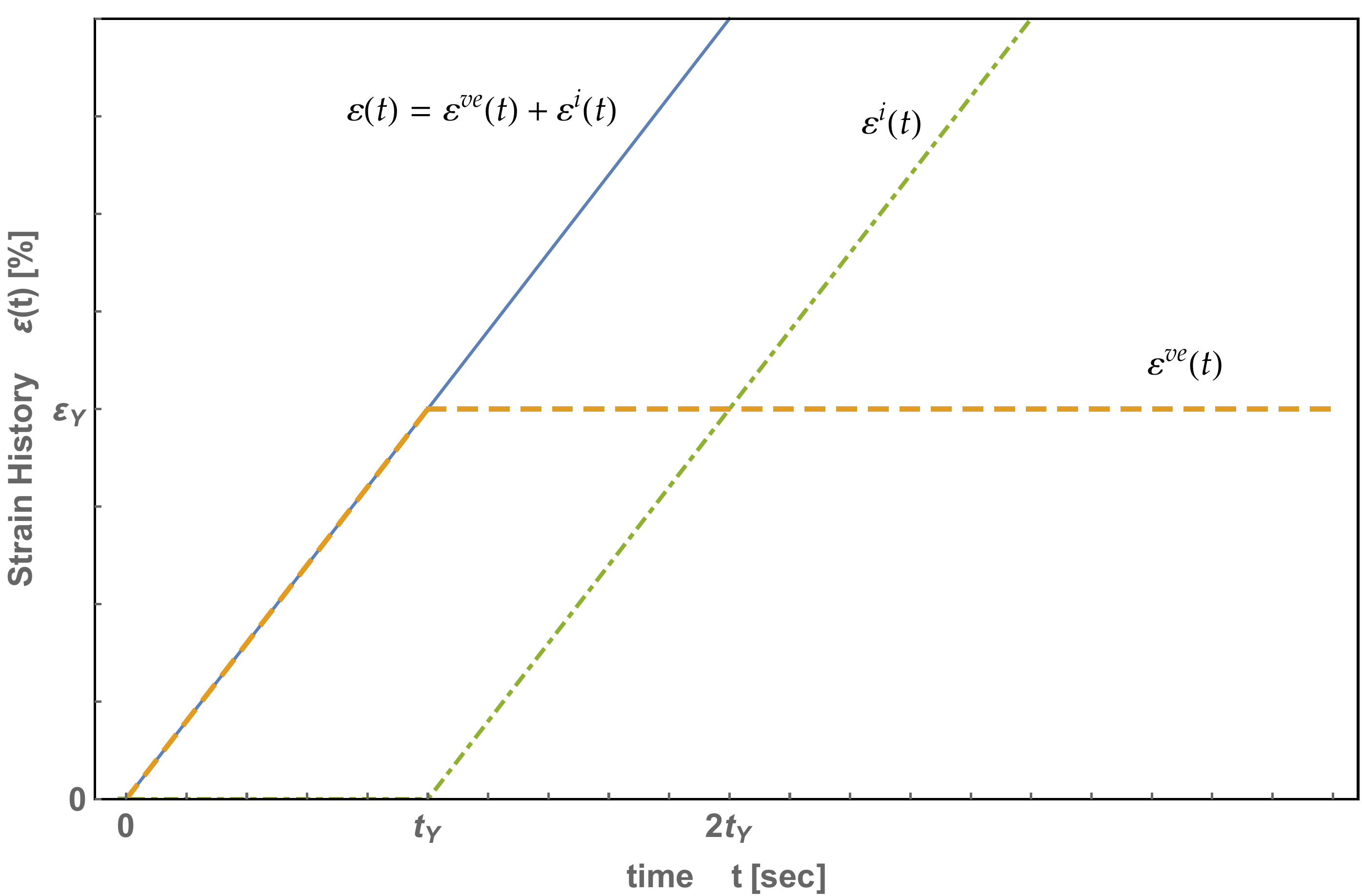}
\caption[]{Imposed strain history $\varepsilon(t)$, viscoelastic $\varepsilon^{ve}(t)$ and inelastic $\varepsilon^i(t)$ deformation.}
\label{ramp_defo}
\end{center}
\end{figure}
Taking into account the assumptions in Eq.s~(\ref{eq6b}), (\ref{eq6}) and (\ref{eq6c}), \figurename~\ref{ramp_defo} shows the imposed deformation history, the viscoelastic strain and the inelastic one. Usually, the limit value of the viscoelastic deformation $\varepsilon_Y$ is a function of the yield stress, $\varepsilon_Y=f(\sigma_Y)$. 
Moreover, according to the Eq.~(\ref{eq16}) and under the aforementioned assumptions, the viscoelastic and inelastic deformation are
\begin{subequations}
\begin{equation}
\varepsilon^{ve}(t)=\left\{\begin{split}
&0,	&				&t<0,&\\
&\dot\varepsilon_0\,t,	&	&0\leqslant t<t_Y,&\\
&\varepsilon_Y,		&	&t\geqslant t_Y,&
\end{split}
\right.
\end{equation}
\begin{equation}
\varepsilon^i(t)=\left\{\begin{split}
&0,	&				&t<t_Y,&\\
&\dot\varepsilon_0\,(t-t_Y),	&	&t\geqslant t_Y.&\\
\end{split}
\right.\end{equation}
\label{eq18a}
\end{subequations}
\begin{subequations}
For $t\geqslant0$ strain histories in Eq.s~(\ref{eq18a}) may be rewritten as
\begin{equation}
\varepsilon^{ve}(t)=\dot\varepsilon_0\,t\,U(t_Y-t)+\varepsilon_Y\,U(t-t_Y)
\end{equation}
\begin{equation}
\varepsilon^i(t)=\dot\varepsilon_0\,(t-t_Y)\,U(t-t_Y)
\end{equation}
\label{eq18b}
\end{subequations}

\subsection{Stress response for imposed strain history}
By placing the strain histories of Eq.s~(\ref{eq18b}) into the fractional stress-strain relation of Eq.~(\ref{eq10}), the following relation holds true
\begin{equation}
\sigma(t)=\left\{\begin{split}
&\bar{A}(\dot\varepsilon_0)t^{1-\alpha},&    &0<t< t_Y,\\
&\bar{A}(\dot\varepsilon_0)\left[t^{1-\alpha}-(t-t_Y)^{1-\alpha}\right]+\bar{B}(\dot\varepsilon_0)(t-t_Y)^{1-\beta},&    &t\geqslant t_Y,
\end{split}\right.
\end{equation}
or in compact form
\begin{equation}
\sigma(t)=\bar{A}(\dot\varepsilon_0)t^{1-\alpha}-\left[\bar{A}(\dot\varepsilon_0)(t-t_Y)^{1-\alpha}-\bar{B}(\dot\varepsilon_0)(t-t_Y)^{1-\beta}\right]U(t-t_Y),
\label{eq19}
\end{equation}
where the involved coefficients are
\begin{equation}
\bar{A}(\dot\varepsilon_0)=\frac{\tilde{A}\dot\varepsilon_0}{\Gamma(2-\alpha)},\;\;\;\;\;\;\;
\bar{B}(\dot\varepsilon_0)=\frac{\tilde{B}\dot\varepsilon_0}{\Gamma(2-\beta)}.
\end{equation}

\begin{figure}[h]
\begin{center}
\includegraphics[width=0.475\textwidth]{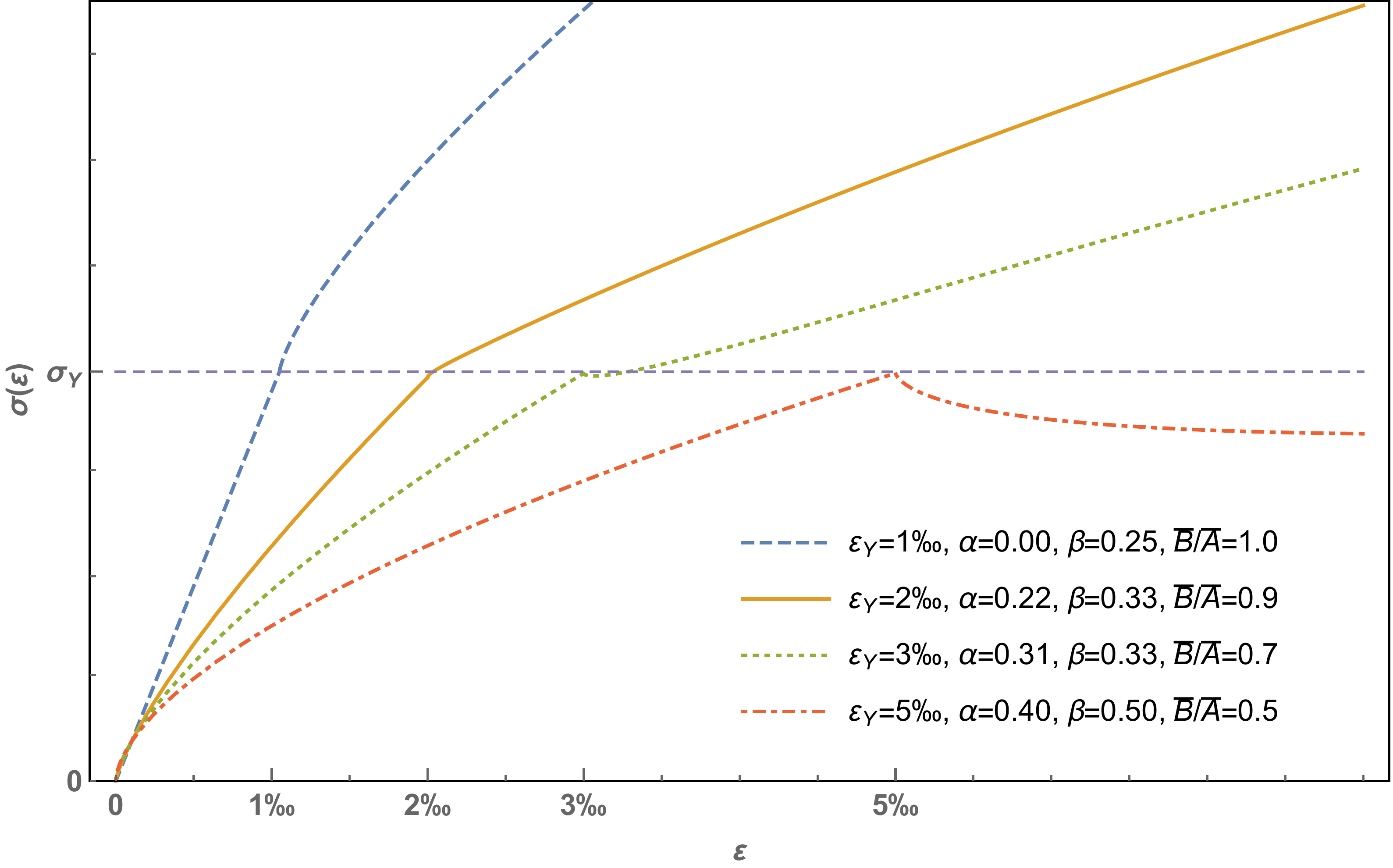}
\caption[]{Stress-strain relation by Eq.~(\ref{eq20}) for different values of the parameters.}
\label{stress-strain}
\end{center}
\end{figure}

Moreover, taking into account the imposed strain history in Eq.~(\ref{eq16}) and performing a change of the variable from $t$ to $\varepsilon$, the relation in Eq.~(\ref{eq19}) can be rewritten as
\begin{equation}
\sigma(\varepsilon)=\bar{A}(\dot\varepsilon_0)\varepsilon^{1-\alpha}-\left[\bar{A}(\dot\varepsilon_0)(\varepsilon-\varepsilon_Y)^{1-\alpha}-\bar{B}(\dot\varepsilon_0)(\varepsilon-\varepsilon_Y)^{1-\beta}\right]U(\varepsilon-\varepsilon_Y).
\label{eq20}
\end{equation}
The five involved parameters $\tilde A$, $\alpha$, $\tilde B$, $\beta$ and $\varepsilon_Y$ have to be evaluated by performing a best-fitting of experimental data. \figurename~\ref{stress-strain} shows some stress-strain relations obtained by Eq.~(\ref{eq20}) with different values of the involved parameters.

\subsection{Idealized stress-strain curves from the proposed model}
The proposed stress-strain relation in Eq.~(\ref{eq20}) represents a rate-dependent non-linear constitutive law describing the evolution of the stress during a displacement-control tensile tests. The model needs the definition of five parameters, that is, two coefficients $A$ and $B$ (anomalous moduli), two related fractional orders $\alpha$ and $\beta$, and a yielding value $\varepsilon_Y$. 

\begin{figure}[h]
\begin{center}
\includegraphics[width=0.475\textwidth]{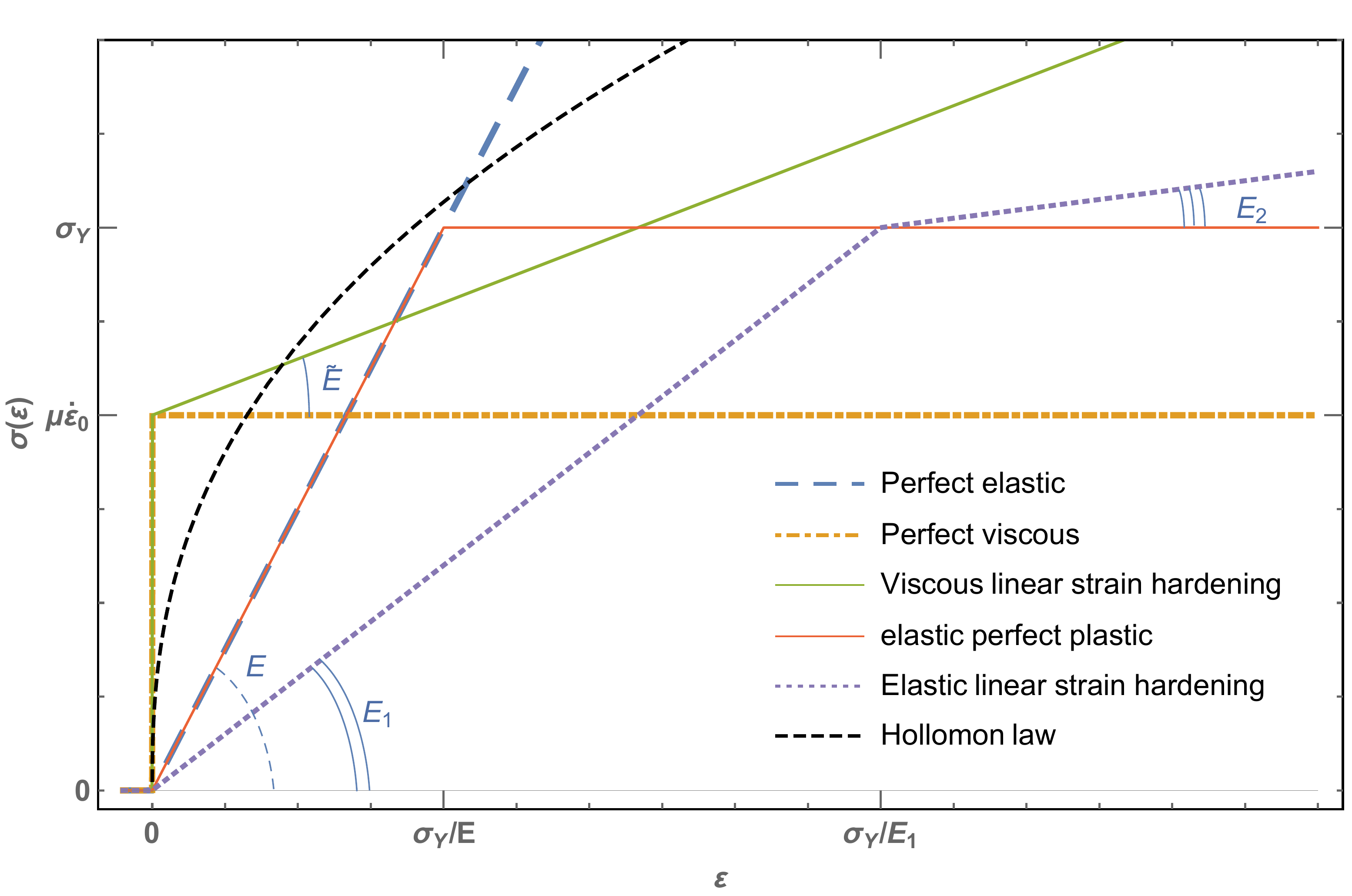}
\caption[]{Some known cases obtained as particular case of the Eq.~(\ref{eq20}).}
\label{knowncases}
\end{center}
\end{figure}

From the Eq.~(\ref{eq20}), with a proper selection of the involved five parameters, some the particular known cases reported in \figurename~\ref{knowncases} can be derived. In particular, 
\begin{itemize}
\item if the yielding strain is  such that $\varepsilon_Y\gg0$, and the fractional order $\alpha=0$, then the anomalous modulus $A$ becomes the classical Young modulus $A=E$, and the \emph{perfect elastic} case is obtained; 
\item if the yielding strain is  such that $\varepsilon_Y\gg0$, and the fractional order $\alpha=1$, then the anomalous modulus $A$ becomes the Newtonian viscosity $\mu$, and the proposed stress-strain relation leads to the \emph{perfect viscous} model;
\item if the yielding strain is  such that $\varepsilon_Y=0$, $\alpha=1\Rightarrow A=\mu$, and $\beta=0\Rightarrow B=\tilde E$, a \emph{viscous linear strain hardening} behavior is obtained;
\item if the yielding strain is  such that $\varepsilon_Y=\sigma_Y/A$, and $\alpha=0\Rightarrow A=E\gg B$, the \emph{elastic perfect plastic} case is derived;
\item if the yielding strain is  such that $\varepsilon_Y=\sigma_Y/A$, $\alpha=0\Rightarrow A=E_1$, and $\beta=0\Rightarrow B=E_2$ another particular case is obtained, that is, the \emph{elastic linear strain hardening}, where $E_1$ and $E_2$ are the stiffness before and after the yielding, respectively;
\item if the yielding strain is zero, $\varepsilon_Y=0$, $\beta=n_H$, and $B=K\gg A$, according to Eq.~(\ref{eq1}), fractional stress-strain relation becomes the \emph{Hollomon law}.
\end{itemize}

\section{Best fitting of experimental data} \label{sect4} 
\noindent The proposed stress-strain relation is used here to obtain a best-fitting of some experimental data. The considered experiments are reported in \cite{exp3,Aginagalde,Perez}. Such experiments regard a tensile test on two metals: \emph{AA60820-O} aluminum alloy and \emph{AHSS TRIP 700 steel}. Both tests are performed at room temperature with imposed strain ratio $\dot\varepsilon_0=0.001$ $s^{-1}$. 

In order to show the capabilities of the proposed model, the best-fitting of experimental data obtained with the Eq.~(\ref{eq20}) is compared to the ones obtained with other known models. In particular, the classical models of Hollomon in Eq.~(\ref{eq1}) and Ramberg-Osgood in Eq.~(\ref{eq2}) are considered. Moreover, a recent rate-independent model based on fractional calculus, proposed by Mendiguren et al. in \cite{exp3}, is also considered. Such model is composed by two fractional terms and it needs the determination of four parameters. In particular, the stress-strain relation in \cite{exp3} is obtained from the following fractional differential equation
\begin{equation}
a_1\left(D^{\alpha_1}\sigma\right)(\varepsilon)+a_1\left(D^{\alpha_1}\sigma\right)(\varepsilon)=1.
\label{mendi2}
\end{equation}
where the $\left(D^{\alpha_j}\sigma\right)(\varepsilon)$ denotes the $\alpha_j$-order Grunwald-Letinokov fractional derivative of the stress respect to the strain \cite{Oldham,Samko,Miller,Pod,Kilbas}. That is,  
\begin{equation}
\left(D^{\alpha_j}\sigma\right)(\varepsilon)=\lim_{\Delta \varepsilon \to 0}\frac{1}{\Delta \varepsilon^{\alpha_j}}\sum_{k=0}^{\frac{t-\Delta \varepsilon}{\Delta \varepsilon}}\frac{\Gamma(k-\alpha_j)}{\Gamma(-\alpha_j)\Gamma(k-1)}\sigma(\varepsilon-j\Delta \varepsilon).
\label{mendiB}
\end{equation} 
The solution of Eq.~(\ref{mendi}) is
\begin{equation}
\sigma(\varepsilon)=\sum_{k=0}^\infty\frac{(-1)^ka_1^k\,\varepsilon^{\alpha_2(k+1)-k\alpha_1}}{a_2^{k+1}\Gamma\left[\alpha_2(k+1)-k\alpha_1+1\right]},
\label{mendi}
\end{equation}
where the parameters $a_1$, $\alpha_1$, $a_2$, and $\alpha_2$ for the considered experiments are detailed in \cite{exp3}.

The parameters resulting from the best fitting procedure of the proposed stress-strain relation in Eq.~(\ref{eq20}) and of the other three benchmark models are reported in Tables~\ref{aluminum} and \ref{steel}. In the first table the parameters of the {AA60820-O} aluminum alloy are reported, whereas the second one contains the parameters related to the {AHSS TRIP 700 steel}. The first three rods of the contain the parameters of the known models that are obtained in \cite{Mendiguren}. The five parameters of the proposed model are reported in the forth rods, they are obtained by least-squares method with the aid of the software \emph{Wolfram Mathematica}.

\begin{table}[h]
  \centering 
  \begin{tabular}{l l l l l l}
\hline\toprule
\textbf{Model}					& \textbf{Parameters }	&  		&	&	&  \\[1mm]
\hline
\toprule\\[-4mm]
Hollomon Eq.~(\ref{eq1})			& $K$ (MPa)	&$n_H$		&	&	&	 \\
							& $235.77$ 	&$0.1812$	&	&	&	  \\
							\hline\\[-3mm]
Ramberg-Osgood Eq.~(\ref{eq2})	& $E$ (MPa)  	&$n_{RO}$  	&$H$ (MPa)	&	&  \\
							& $70000.00$ 	&$5.7452$	&$233.07$	&	&	 \\
							\hline\\[-3mm]
Mendiguren et al. Eq.~(\ref{mendi})	& $a_1$ (MPa$^{-1}$)	&$\alpha_1$ & $a_2$ (MPa$^{-1}$) &$\alpha_2$  & 	  \\
							& $4.6411\times10^{-3}$ 	&$0.1710$		&$1.4286\times10^{-5}$	&$1.0000$ &  \\
							\hline\\[-3mm]
Fractional stress-strain relation Eq.~(\ref{eq20}) 	& $\bar A$ (MPa)  	&$\alpha$  	&$\bar B$ (MPa)	&$\beta$	&$\varepsilon_Y$  \\
										& $24516.6108$ 	&$0.1500$	&$233.72$%47$	
										&$0.7191$	&$0.0011$  \\
\bottomrule
\end{tabular}
  \caption{Parameters of {AA60820-O} aluminum alloy.}\label{aluminum}
\end{table}

\begin{table}[h]
  \centering 
  \begin{tabular}{l l l l l l}
\hline
\toprule
\textbf{Model}					& \textbf{Parameters }	&  		&	&	&  \\[1mm]
\hline\toprule\\[-4mm]
Hollomon Eq.~(\ref{eq1})			& $K$ (MPa)	&$n_H$		&	&	&  \\
							& $1253.90$ 	&$0.2202$	&	&	&  \\
							\hline\\[-3mm]
Ramberg-Osgood Eq.~(\ref{eq2})	& $E$ (MPa)  	&$n_{RO}$  	&$H$ (MPa)	&	&  \\
							& $203000.00$ 	&$4.8267$	&$1230.10$	&	&  \\
							\hline\\[-3mm]
Mendiguren et al. Eq.~(\ref{mendi})	& $a_1$ (MPa$^{-1}$)	&$\alpha_1$ & $a_2$ (MPa$^{-1}$) &$\alpha_2$  	& 	  \\
							& $8.8637\times10^{-4}$ 	&$0.2034$		&$4.92611\times10^{-6}$	&$0.0012$ 	&  \\
							\hline	\\[-3mm]
Fractional stress-strain relation 	Eq.~(\ref{eq20})	& $\bar A$ (MPa)  	&$\alpha$  	&$\bar B$ (MPa)	&$\beta$	&$\varepsilon_Y$  \\
										& $70000.00$ 		&$0.1820$	&$1271.83$	&$0.6365$	&$0.0023$  \\
\bottomrule
\end{tabular}
  \caption{Parameters of AHSS TRIP 700 steel.}\label{steel}
\end{table}

\begin{figure}[h]
\subfigure[{AA60820-O} Aluminum Alloy]{\includegraphics[width=0.475\textwidth]{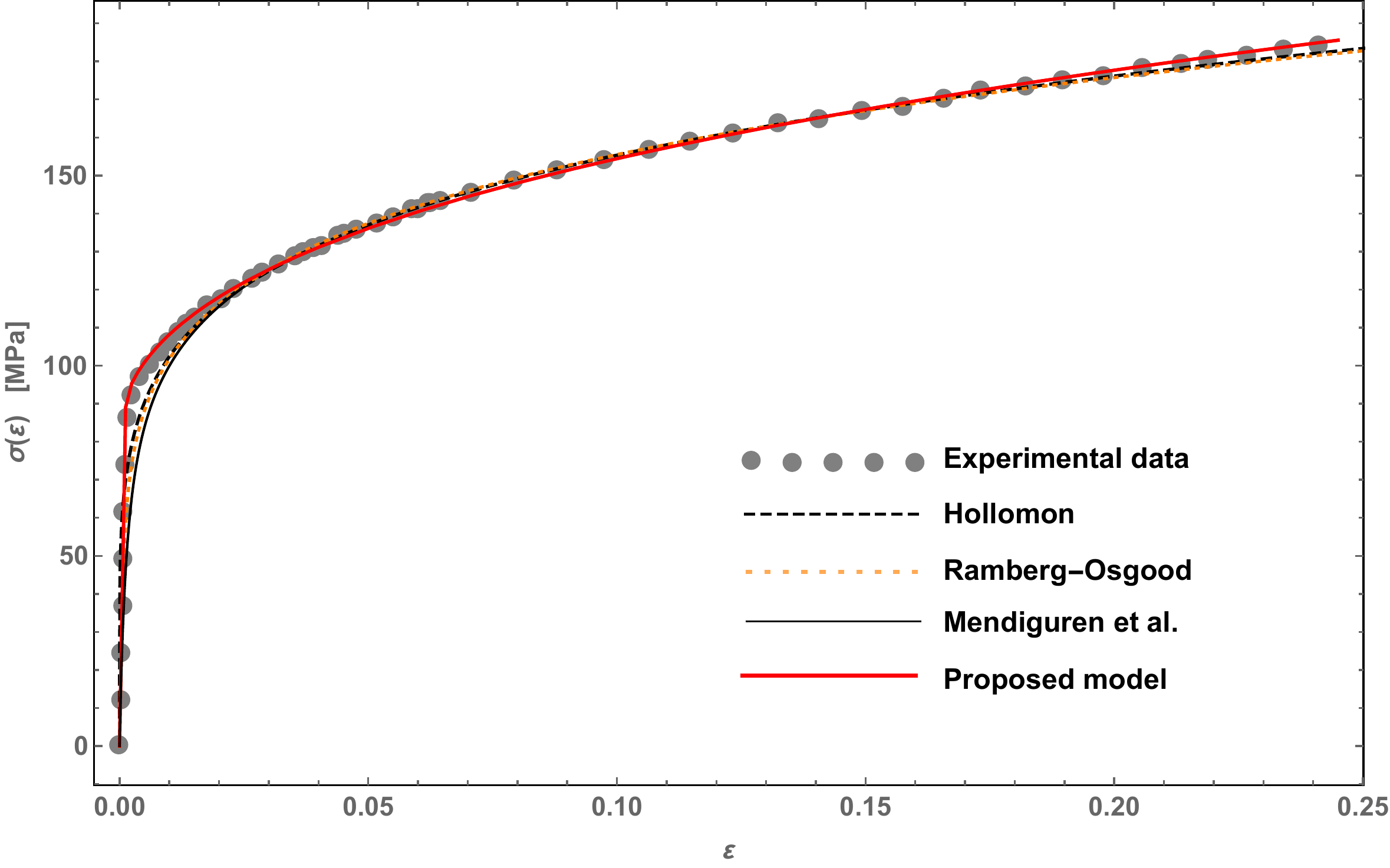}\label{Aluminum entire}}
\subfigure[AHSS TRIP 700 Steel]{\includegraphics[width=0.475\textwidth]{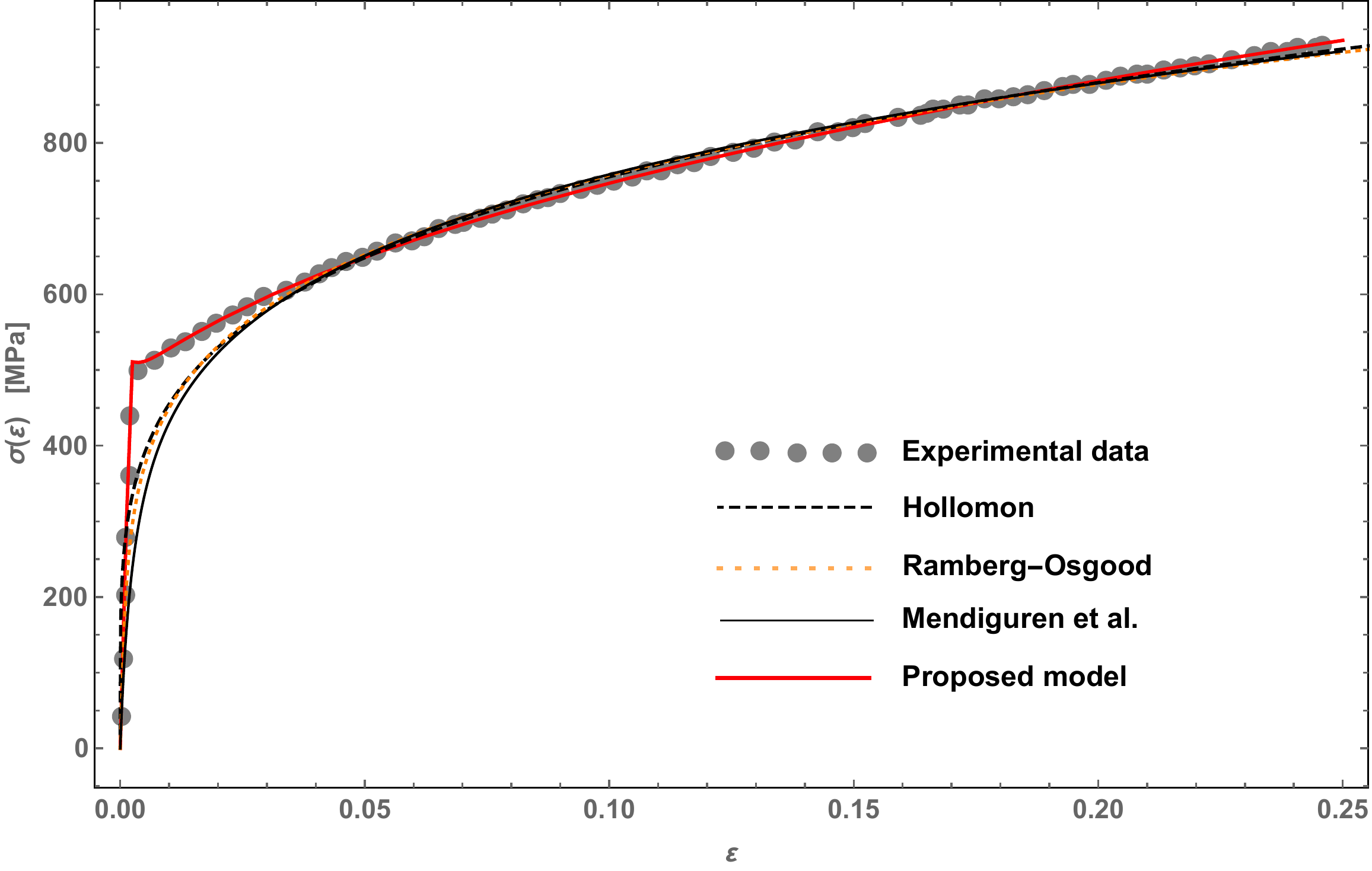}\label{Steel entire}}
\caption{Stress-strain relation: experimental data and four fitted models.}
\label{tensile tests}
\end{figure}

\figurename~\ref{tensile tests} shows the stress-strain relation during the whole tensile test for both the materials. \figurename~\ref{Aluminum entire} shows the experimental data (dotted line) of the {AA60820-O} aluminum alloy, the fitting laws of the proposed model and of other three, using the parameters reported in \tablename~\ref{aluminum}. \figurename~\ref{Steel entire} shows the comparison between experimental data of {AHSS TRIP 700 steel}, the proposed model and the others, for which the parameters are reported in \tablename~\ref{steel}. From these figures it is possible to observe that the best agreements with the considered experimental data are obtained with the proposed fractional stress-strain relation.

\begin{figure}[h]
\subfigure[{AA60820-O} Aluminum Alloy]{\includegraphics[width=0.475\textwidth]{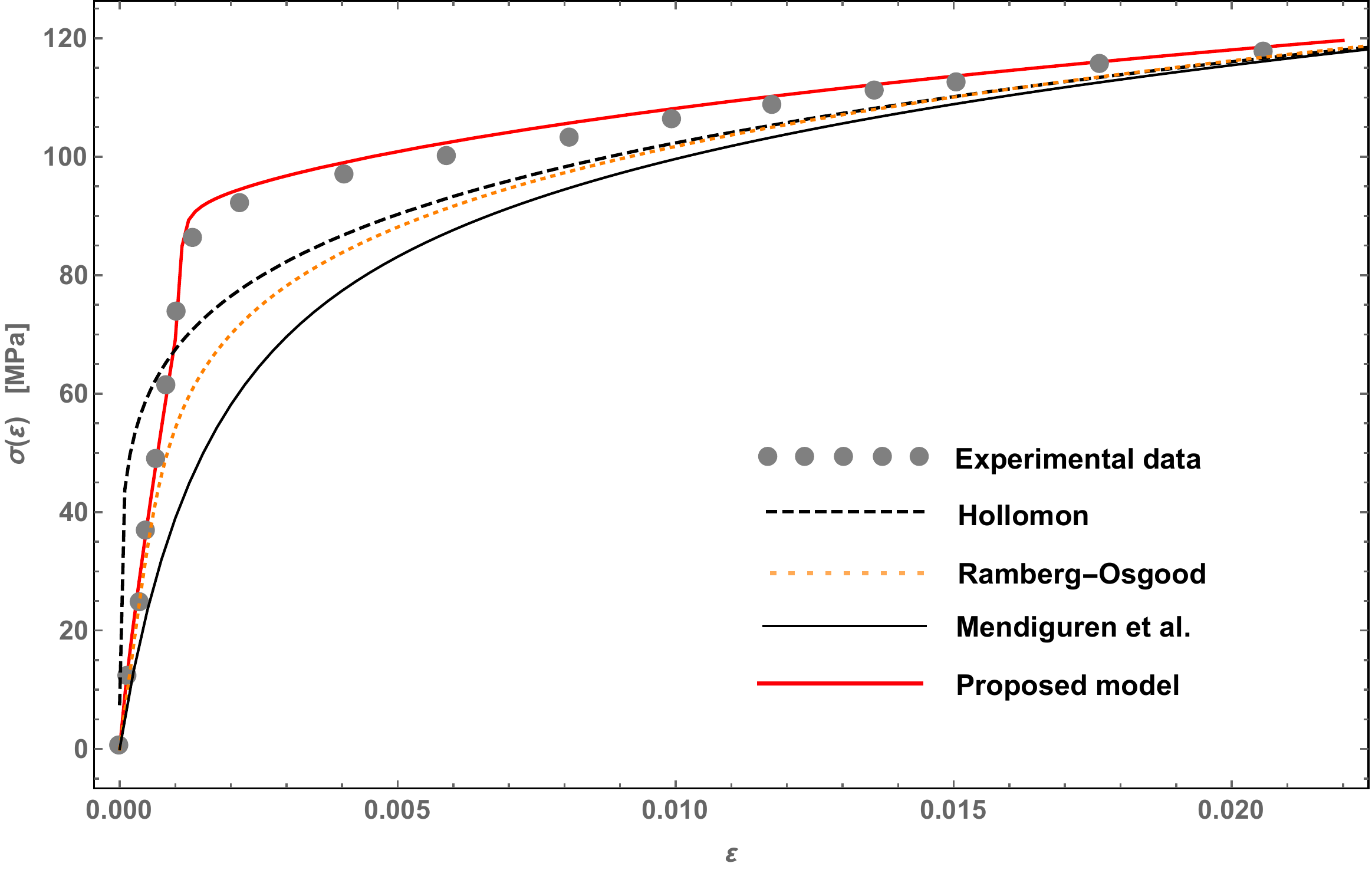}}
\subfigure[AHSS TRIP 700 Steel]{\includegraphics[width=0.475\textwidth]{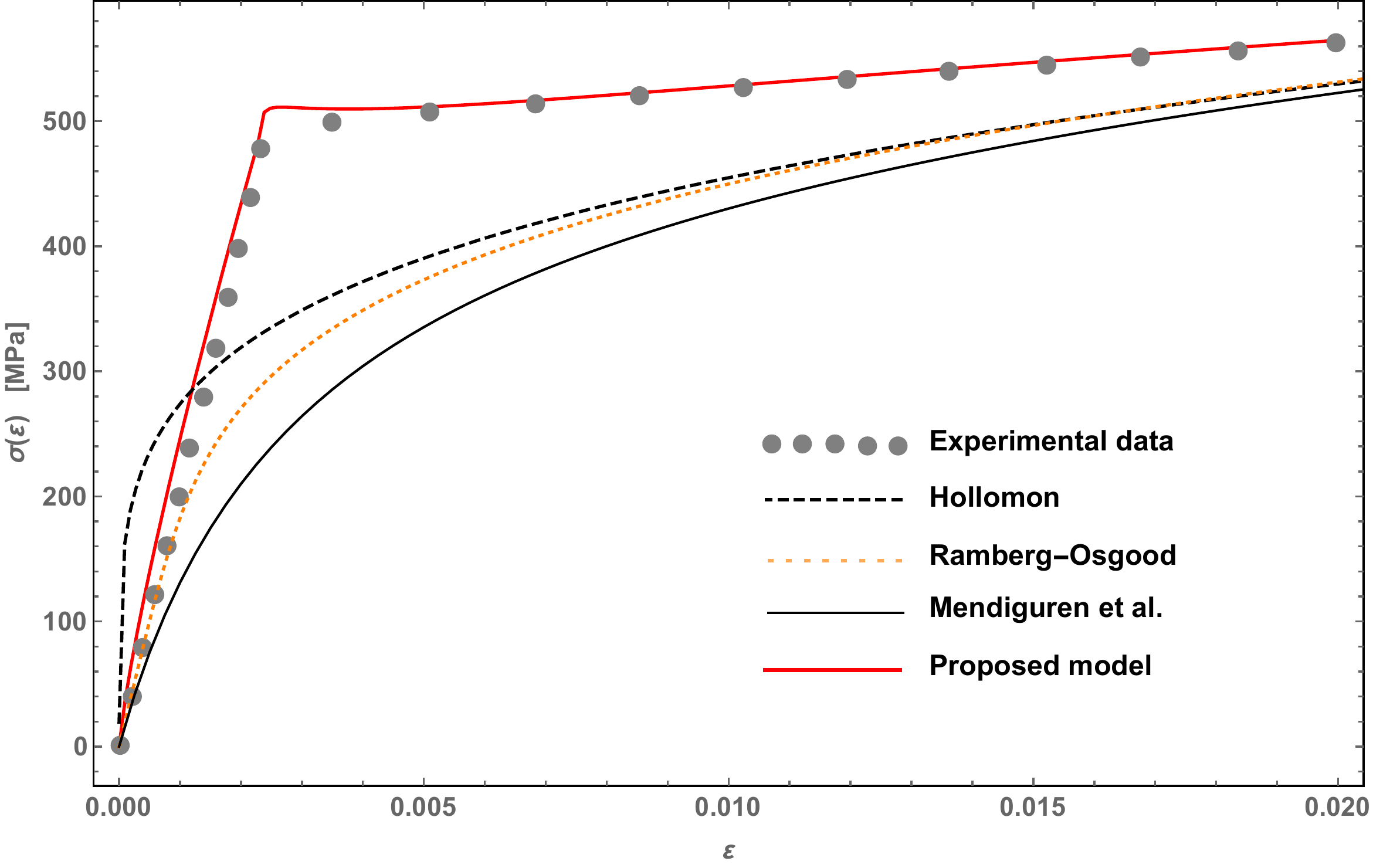}}
\caption{{Enlargement of the} stress-strain relation near the yielding point: experimental data and four fitted models.}
\label{initial tensile tests}
\end{figure}

\begin{figure}[h]
\subfigure[{AA60820-O} Aluminum Alloy]{\includegraphics[width=0.475\textwidth]{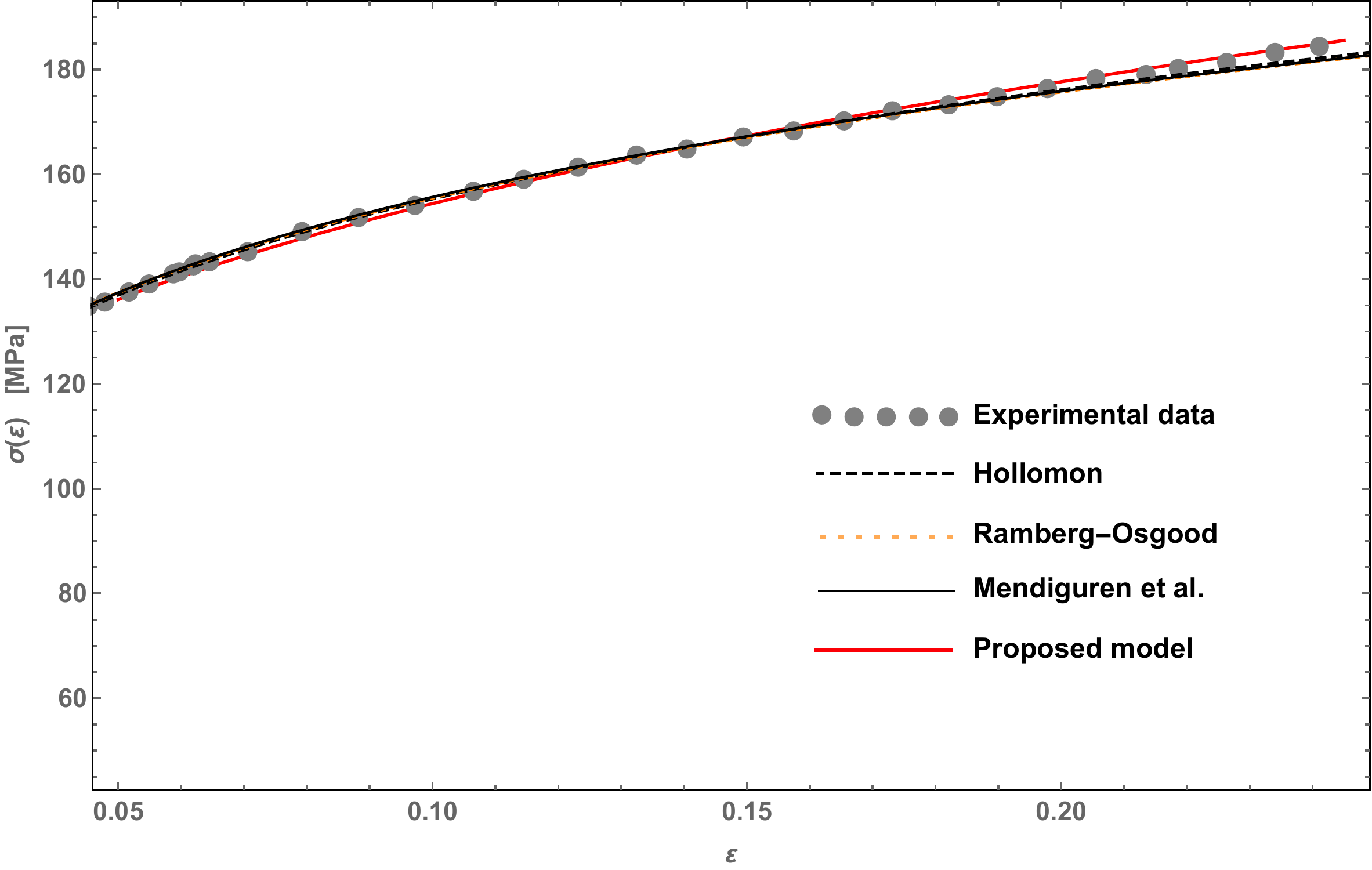}}
\subfigure[AHSS TRIP 700 Steel]{\includegraphics[width=0.475\textwidth]{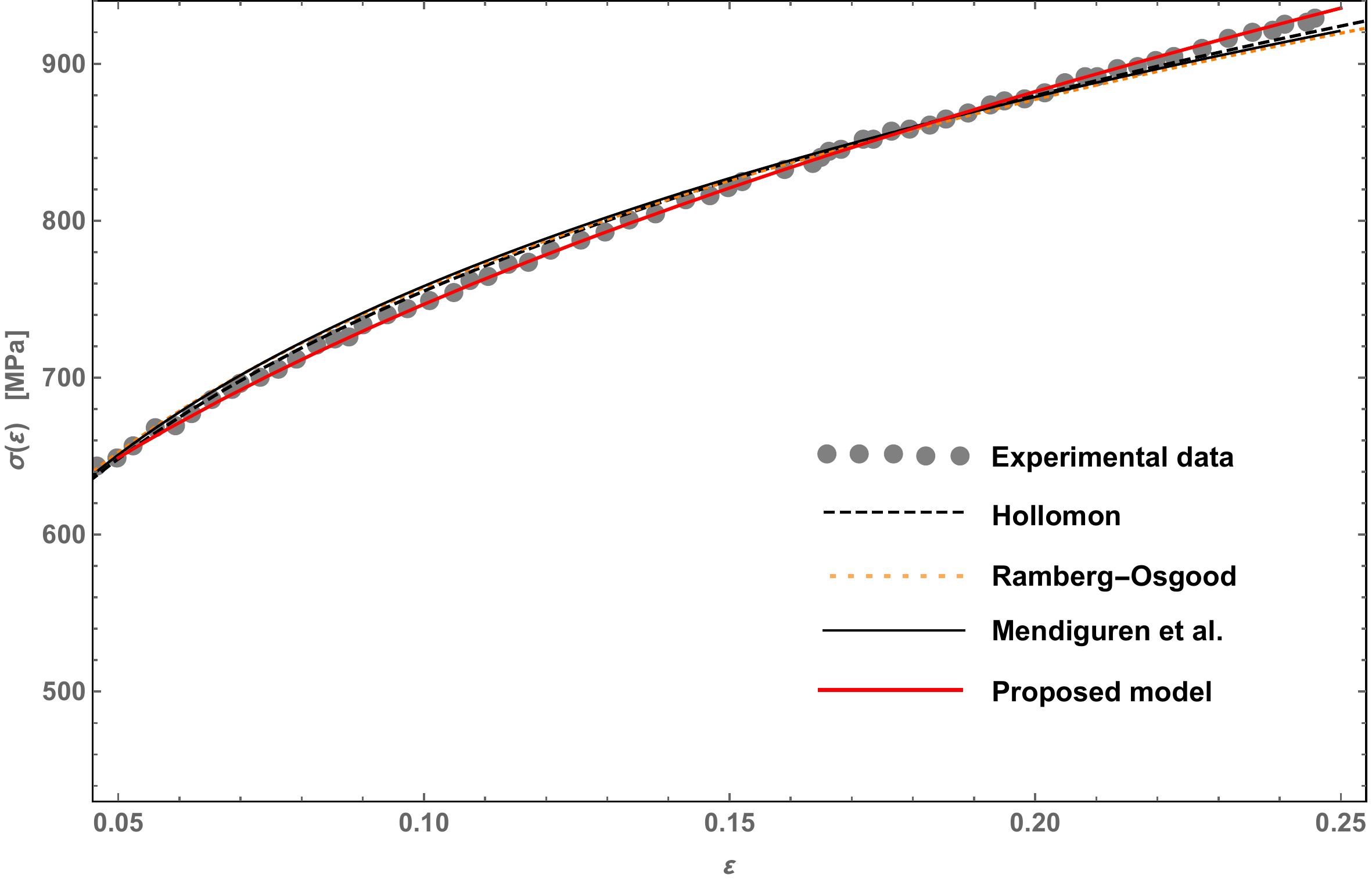}}
\caption{{Enlargement of the} stress-strain relation for large strain: experimental data and four fitted models.}
\label{final tensile tests}
\end{figure}

\figurename s~\ref{initial tensile tests} show the details of the stress-strain curves close the yielding point. From these figures it can be observed that the proposed model is able to fit experimental data with excellent accuracy in this particular zone of the curves. Moreover, the proposed model is also able to fit the experimental results also for large strain. In particular, in \figurename~\ref{final tensile tests} the comparisons between the experimental data and the considered models for large value of strain are reported. Observe that also in this case the proposed model guarantees the best agreement between theoretical and experimental results.

As can be seen from the \figurename s~\ref{tensile tests} and \ref{final tensile tests}, all the models offer a good agreement for greater value of the deformation, but \figurename s~\ref{initial tensile tests} show that only the proposed model is also able to simulate the initial stress-strain relation with a good agreement with the experimental tests. 

For the comparison between the different laws and the evaluation of the accuracy of the models for the best-fitting, two error parameters are evaluated: the \emph{mean square error} ($MSE$) and the \emph{mean absolute percentage error} ($MAPE$). $MSE$ measures the average of the squares of the errors. It is defined as the difference between the exact value and theoretical one obtained by the model. That is,
\begin{equation}
MSE=\frac{1}{n}\sum_{j=1}^n\left[\sigma_j-\sigma(\varepsilon_j)\right]^2,
\label{MSE}
\end{equation}
where $\sigma_j$ is $i$-th experimental value of the stress, and $\sigma(\varepsilon_j)$ is the stress obtained by the considered model for the $i$-the experimental stress, $n$ are the considered experimental values. The other error parameter, $MAPE$, provides an evaluation of the quality of the considered models in the estimation. It usually expresses accuracy as a percentage by the following expression
\begin{equation}
MAPE=\frac{100\,\%}{n}\sum_{j=1}^n\left|\frac{\sigma_j-\sigma(\varepsilon_j)}{\sigma_j}\right|.
\end{equation}

Both parameters are reported for all the considered models in Table~\ref{errorsAA} and \ref{errorsTRIP}. In particular, in Table~\ref{errorsAA} the $MSEs$ and $MAPEs$ related to the experimental data of the {{AA60820-O} aluminum alloy} are reported, while Table~\ref{errorsTRIP} shows the error parameters of best fitting procedure for the {AHSS TRIP 700 steel} data.
\begin{table}[h]
  \centering 
  \begin{tabular}{l r r}
\hline
\toprule
\textbf{Model}					& $\boldsymbol{MSE}$&  		$\boldsymbol{MAPE}$ \\
\hline\toprule\\[-4mm]
Hollomon Eq.~(\ref{eq1})			& $46.7713$	&$8.89\,\%$	  \\
							\hline\\[-3mm]
Ramberg-Osgood Eq.~(\ref{eq2})	& $45.1300$  	&$5.34\,\%$  	 \\
							\hline\\[-3mm]
Mendiguren et al. Eq.~(\ref{mendi})	& $112.6147$	&$7.98\,\%$   \\
							\hline	\\[-3mm]
Fractional stress-strain relation 	Eq.~(\ref{eq20})	& $4.5657$  	&$2.61\,\%$\\
\bottomrule
\end{tabular}
  \caption{MSEs and MREs  of {{AA60820-O} aluminum alloy} data best-fitting.}\label{errorsAA}
\end{table}
\begin{table}[h]
  \centering 
  \begin{tabular}{l r r}
\hline
\toprule
\textbf{Model}					& $\boldsymbol{MSE}$&  		$\boldsymbol{MAPE}$ \\
\hline\toprule\\[-4mm]
Hollomon Eq.~(\ref{eq1})			& $1061.7008$	&$4.47\,\%$	  \\
							\hline\\[-3mm]
Ramberg-Osgood Eq.~(\ref{eq2})	& $1165.0348$  	&$3.37\,\%$  	 \\
							\hline\\[-3mm]
Mendiguren et al. Eq.~(\ref{mendi})	& $997.2121$	&$5.01\,\%$   \\
							\hline	\\[-3mm]
Fractional stress-strain relation 	Eq.~(\ref{eq20})	& $160.1546$  	&$1.84\,\%$\\
\bottomrule
\end{tabular}
  \caption{MSEs and MREs of {AHSS TRIP 700 steel} data best-fitting.}\label{errorsTRIP}
\end{table}
From such tables, it is possible to observe that the lowest value of the error is obtained by the stress-strain relation in Eq.~(\ref{eq20}). Therefore, the best agreement with the experimental data is obtained by the proposed mechanical model. 

The proposed stress-strain relation is able to fit the mechanical behavior before and after the yielding point. Compared to the other models, the proposed model is able to better describe the mechanical behavior near the yielding point. Such major capability to fit experimental data is related to three fundamental aspects of the proposed model. In particular, i) viscoelastic stress-strain relation is represented by Boltzmann superposition integral with power-law kernel; ii) the yielding phenomenon activates an inelastic behavior; iii) the inelastic stress-strain relation is modeled by an integral formulation with non-linear kernel. 

The proposed stress-strain relation is non-linear after the yielding phenomenon, however for particular cases it becomes just the summation of two fractional derivatives with different orders. The appearance of these two linear operators simplified the stress-strain relation. Therefore, although the model requires the evaluation of five parameters, for the considered cases, the evaluation of the mechanical parameters is easy.

\section{Concluding remarks} \label{sect5}
\noindent Considering the evidence presented above, it is evident that the proposed stress-strain relation is able to predict accurately the mechanical behavior of metals during the tensile test. Such stress-strain relation may take into account both viscoelastic and plastic behaviors. It has been obtained as an integral formulation of the constitutive law. The model is similar to the endochronic theory of plasticity introduced by Valanis. The main difference between the proposed model and the Valanis' theory lies in two aspects, that is, the definition of the yielding surface and the use of a time power-law kernel in the convolution integrals.

The proposed model provides a non-linear formulation of the stress-strain relation. However, if the strain history is a monotonic increasing function of time, the stress history becomes a summation of two time fractional derivatives of the strain history. The first one describes the viscoelastic stress, while the second one is related to the mechanical behavior after the yielding phenomenon.

Considering that the most used test for the characterization of the mechanical properties of the materials is the tensile one, the stress-strain relation has been particularized for the case in which the imposed strain history is a ramp function in time. It is shown that, for the tensile test, the stress-strain relation becomes a summation of power-law functions with five parameters. Such parameters have been evaluated by a best-fitting procedure for two metal alloys. After the parameter evaluation, the results of the proposed model have been compared to other ones obtained from other known models. Such comparison has shown that the proposed model offers the best agreement with the experimental data and the lowest level of the error.

\section*{Acknowledgements}
\noindent Francesco P. Pinnola and Giorgio Zavarise gratefully acknowledge the support received from the Italian Ministry of University and Research, through the PRIN 2015 funding scheme (project 2015JW9NJT \emph{Advanced mechanical modeling of new materials and structures for the solution of 2020 Horizon challenges}). Antonio Del Prete and Rodolfo Franchi gratefully acknowledge the support received from the Italian Ministry of University and Research, through the P.O.N. RICERCA E COMPETITIVIT\'A 2007-2013 funding scheme (project PON03PE 00067 \emph{TEMA TEchnologies for Production and Maintenance applied to Aeronautic Propulsion}).

%\section*{References}

\label{lastpage}

\end{document}